\documentclass[12pt]{article}
\usepackage{graphicx}
\usepackage{amsmath, amssymb,bbm}
\usepackage{hfstyle}

\newtheorem{theorem}{Theorem}

\newcommand{\NN}{{\mathbbm{N}}}
\newcommand{\RR}{{\mathbbm{R}}}

\begin{document}
\author{Henryk Fuk\'s
      \oneaddress{
         Department of Mathematics\\
         Brock University\\
         St. Catharines, Ontario  L2S 3A1, Canada\\
        \texttt{hfuks@brocku.ca}}}
\title{Second order additive invariants in elementary cellular automata}
\Abstract{
We investigate second order additive invariants in elementary cellular automata rules.  
Fundamental diagrams of rules which possess additive invariants are either linear or exhibit
singularities similar to singularities of rules with first-order invariant. Only rules which have exactly one
invariants exhibit singularities. At the singularity, the current decays to its equilibrium value as a  power law $t^{\alpha}$,
and  the value of the exponent $\alpha$ obtained from numerical simulations is very close
to $-1/2$. This is in agreements with values previously reported for number-conserving rules,
and leads to a conjecture that  regardless of the order of the invariant,
exponent $\alpha$ seems to have a universal value of $1/2$.
} \maketitle

\section{Introduction}
Cellular automata (CA) are often described as systems of cells in a regular lattice
updated synchronously according to a local interaction rule.
An interesting subclass of CA consists of rules possessing an additive
invariant. The simplest of such invariants is the total number of
sites in a particular state. CA with such invariant, often called
``conservative CA'' or ``number-conserving CA'',
 generated a lot of interest in recent years
\cite{Pivato02,Moreira03,Durand2003,Formenti2003,Morita2001}. 
Number-conserving  CA can be viewed as a
system of interacting and moving particles, where in the case of a
binary rule, 1's represent sites occupied by particles, and 0's
represent empty sites. The flux or current of particles in equilibrium
depends only on their density, which is invariant. The graph of the
current as a function of density characterizes many features of the
flow, and is therefore called the fundamental diagram. 

For a majority of number-conserving  CA rules, fundamental diagrams are piecewise-linear,
usually possessing one or more ``sharp corners'' or singularities.
There exist a strong evidence of universal behavior at singularities,
as reported in \cite{paper19,paper25}.

Since number-conserving CA are simplest rules with additive invariants, 
it would be interesting to consider higher order invariants and CA rules
with such invariants. In 1991, Hattori and Takesue performed
extensive study of additive invariants  in discrete-time lattice
dynamical systems \cite{Hattori91}, not necessarily restricted to CA. They derived
very general existence conditions, and applied them to elementary CA
rules as well as reversible CA. In this paper, we will use their results to study
second-order invariants in elementary rules, focusing mainly on
fundamental diagrams.

\section{Number-conserving cellular automata}

In what follows, we will assume that the dynamics takes place on
one-dimensional lattice of length $L$ with periodic boundary
conditions. Let $s_i(t)$ denote the state of the lattice site $i$ at
time $t$, where $i\in \{ 0, 1, \ldots, L - 1 \}$, $t \in \mathbbm{N}$.
All operations on spatial indices $i$ are assumed to be modulo $L$. We
will further assume that $s_i(t) \in \{ 0, 1\}$, and we will say that
the site $i$ is occupied (empty) at time $t$ if $s_i ( t ) = 1$ ($s_i
( t ) = 0$).

Let $l$ and $r$ be two integers such that $l \leq 0 \leq r$, and let
$n =r-l+1$. The set $\{ s_{i + l} ( t ), s_{i + l + 1} ( t ), \ldots,
s_{i + r} ( t )\}$ will be called the \textit{neighbourhood} of the
site $s_i ( t )$. Let $f$ be a function $f : \{ 0, 1 \}^n \rightarrow
\{ 0, 1 \}$, also called a \emph{local function} The update rule for
the cellular automaton is given by
\begin{equation} 
  \label{cadef} s_i ( t + 1 )_{} = f ( s_{i + l} ( t ), s_{i + l + 1} ( t ),
  \ldots, s_{i + r} ( t ) ) .
\end{equation}

In \cite{Hattori91},  the concept of additive invariant for CA
has been introduced. Let $\alpha$ be a non-negative integer, and let 
$\xi=\xi(x_0,x_1,\ldots, x_{\alpha})$ be a function of $\alpha +1$
variables taking values in~$\RR$. We say that $\xi$
is a density function of an additive conserved quantity
if for every positive integer $L$ and for every
initial condition $(s_0(0),s_1(0),\ldots,s_{L-1}(0)) \in \{ 0, 1 \}^L$
we have
\begin{equation} \label{defadditive}
\sum_{i=0}^{L-1} \xi(s_i(t),s_{i+1}(t),\ldots, s_{i+\alpha}(t))= 
\sum_{i=0}^{L-1} \xi(s_i(t+1),s_{i+1}(t+1),\ldots, s_{i+\alpha}(t+1))
\end{equation} 
for all $t \in \NN$.
For simplicity, if the above condition is satisfied, we will say that $\xi$ 
is an additive invariant of $f$. It is often more convenient to write (\ref{defadditive})
using the function $G$ defined as
\begin{equation}
G(x_0,x_1,\ldots,x_{\alpha+n-1})=
\xi(f(x_0,x_1,\ldots,x_{n-1}), f(x_1,x_2,\ldots,x_{n}),\ldots,
f(x_\alpha,x_{\alpha+1},\ldots,x_{\alpha + n-1})).
\end{equation} 
With this notation, $\xi$ is an additive invariant of $f$ if 
\begin{equation}
\sum_{i=0}^{L-1} G(x_i,x_{i+1},\ldots,x_{i+\alpha+n-1})=
\sum_{i=0}^{L-1} \xi(x_i,x_{i+1},\ldots, x_{i+\alpha})
\end{equation} 
for every positive integer $L$ and for all $x_0,x_1,\ldots, x_{L-1} \in \{0,1\}$.

In recent years, many authors studied the case of the simplest additive invariant,
with $\alpha=0$ and $\xi(x_0)=x_0$. For this invariant, the equation (\ref{defadditive})
becomes 
 \begin{equation} 
\sum_{i=0}^{L-1} s_i(t)= 
\sum_{i=0}^{L-1} s_i(t+1),
\end{equation} 
which means that the CA rule posessing this invariant conserves the number
of sites in state~1. Such rules are often referred to as number-conserving
rules. Among elementary CA, i.e. those with $l=-1$ and $r=1$, there are only
five number-conserving rules. Three of these are trivial, namely the identity
rule 204 and two shifts 170 and 240. Two remaining rules, 184 and 226, are equivalent under the 
spatial reflection. Rule 184, which is a discrete version
of the totally asymmetric exclusion process, has been extensively studied
\cite{Krug88,Nagatani95,Nagel96,Belitsky98,paper11,NishinariT98,BelitskyKNS01,Blank03},
and many rigorous result regarding its dynamics have been established.

Hattori and Takesue \cite{Hattori91} established a very general
result which we will write here in a somewhat simplified form, taking into account that 
this paper is concerned with binary rules only.

\begin{theorem}[Hattori \& Takesue '91]
Let $\xi(x_0,x_1,\ldots,x_\alpha)$ be a function of $\alpha + 1$ variables.
Then $\xi$ is a density function of an additive conserved quantity
under the time evolution of cellular automaton rule (\ref{cadef})
if and only if the condition
\begin{equation} \label{gencondition}
G(x_0,x_1,\ldots,x_{\alpha+n-1})-\xi(x_{-l},x_{-l+1},\ldots,x_{\alpha-l})
=J(x_0,x_1,\ldots,x_{\alpha+n-2})
-J(x_1,x_2,\ldots,x_{\alpha+n-1})
\end{equation} 
holds for all $x_0$, $x_1$, $\ldots$, $x_{\alpha+n-1} \in \{0,1\}$, where
the quantity $J$, to be referred to as the \emph{current}, is defined by
\begin{equation} \label{gencurrentdef}
J(x_0,x_1,\ldots,x_{\alpha+n-2})=
-\sum_{i=0}^{\alpha+n-2} G(\overbrace{0,0,\ldots,0}^{\alpha+n-1-i},x_0,x_1,\ldots,x_i)
+\sum_{i=-l-n+2}^{\alpha-l} \xi(\overbrace{0,0,\ldots,0}^{\alpha+1-i},x_0,x_1,\ldots,x_{i-1}).
\end{equation} 
\end{theorem}

The following convention is used in the definition of $J$:
\begin{equation}
\xi(\overbrace{0,0,\ldots,0}^{\alpha+1-i},x_0,x_1,\ldots,x_{i-1})=
\xi(x_{i-\alpha-1},x_{i-\alpha},\ldots,x_{i-1}) \mbox{\,\,\,\,\,if $i \geq \alpha+1$},
\end{equation} 
and
\begin{equation}
\xi(\overbrace{0,0,\ldots,0}^{\alpha+1-i},x_0,x_1,\ldots,x_{i-1})=\xi(0,0,\ldots,0) 
\mbox{\,\,\,\,\,if $i \leq 0$}.
\end{equation}

The equation (\ref{gencondition}) can be interpreted in a similar way as a
conservation law in a continuous, one dimensional physical system. In
such system, let $\rho(x,t)$ denote the density of some material at
point $x$ and time $t$, and let $j(x,t)$ be the current (flux) of this
material at point $x$ and time $t$. A conservation law states that the
rate of change of the total amount of material contained in a fixed
domain is equal to the flux of that material across the surface of the
domain. The differential form of this condition can be written as
\begin{equation}\label{concons}
    \frac{\partial \rho}{\partial t}=-\frac{\partial j}{\partial
    x}.
\end{equation}
Since in our case $\xi$ is the the density of an additive conserved quantity, the left hand side of
(\ref{gencondition}) is simply the change of density in a
single time step, so that (\ref{gencondition}) is an obvious
discrete analog of the current conservation law (\ref{concons})
with $J$ playing the role of the current.

Let us now assume that the initial configuration has been generated
from some translation-invariant distribution $\mu$. We define the expected value of $\xi$ at site $i$ as
 \begin{equation}
 \rho(i,t)=E_\mu \left[ \xi(s_i(t),s_{i+1}(t),\ldots,s_{i+\alpha}(t)) \right].
 \end{equation}
Since the initial distribution is $i$-independent, we expect that
$\rho(i,t)$ also does not depend on $i$, and we will therefore define
$\rho(t)=\rho(i,t)$.  Furthermore, since $\xi$ is density function of a conserved
quantity, $\rho(t)$ is $t$-independent, so we define $\rho=\rho(t)$. 
The expected value of the
current $J(s_{i+l}(t),s_{i+l+1}(t),\ldots,s_{i+r-1}(t))$ will also be
$i$-independent, so we can define the expected current as
\begin{equation} \label{defexpcurrent}
j(\rho,t)=E_\mu
\big(J(s_{i+l}(t),s_{i+l+1}(t),\ldots,s_{i+r-1}(t))\big).
\end{equation}
 
The graph of the equilibrium current $j(\rho,\infty)=\lim_{t
   \rightarrow \infty} j(\rho,t)$ versus the density $\rho$ is known as
 the fundamental diagram.

\section{Number-conserving nearest-neighbour rules}
In order to illustrate the theorem of the previous section, we will first consider
the case of number-conserving nearest-neighbour rules,  i.e., $\alpha=0$ and $\xi(x_0)=x_0$,
$l=-1$, $n=3$. Condition (\ref{gencondition}) becomes
\begin{equation}
 G ( x_0, x_1, x_2) - \xi(x_{1}) = J ( x_0, x_1)- J (x_1, x_2),
 \end{equation}
where
\begin{equation}
     J(x_0,x_1)=
%     - \sum_{i=0}^{1}
%G(\underbrace{0,0,\ldots,0}_{2-i},x_0,x_1,\ldots,x_{i}) +
%\sum_{i=0}^{1} \xi(\underbrace{0,0,\ldots,0}_{1-i},x_0,x_1,\ldots,x_{i-1})
=-G(0,0,x_0)-G(0,x_0,x_1) + \xi(0)+\xi(x_0)
\end{equation}
Obviously, $G(x_0,x_1,x_2)=f(x_0,x_1,x_2)$, thus the current becomes
$J(x_0,x_1)=-f(0,0,x_0)-f(0,x_0,x_1) - x_0$, and the conservation condition
takes the form
\begin{equation} \label{nnsimpleconscond}
 f( x_0, x_1, x_2) - x_{1} = J ( x_0, x_1)- J (x_1, x_2).
 \end{equation}
As mentioned earlier, rule 184 and its spatial reflection are the only non-trivial
elementary CA rules satisfying (\ref{nnsimpleconscond}). For rule 184, $f$ is defined by
$f(x_0,x_1,x_2)=x_0-x_0 x_1 +x_1 x_2$
and the current can be written as $J(x_0,x_1)=x_0 (1-x_1)$. 
It is possible to show \cite{paper11} that the equilibrium current for this rule is given by 
\begin{equation} 
j(\rho,\infty)= \left\{ \begin{array}{ll}
  \rho  & \mbox{if $\rho<1/2$}, \\
 1-\rho    & \mbox{otherwise}.
\end{array}
\right.
\end{equation}

Since number-conserving CA rules conserve the number of occupied
sites, we can label each occupied site (or ``particle'') with an
integer $k\in\mathbbm{Z}$, such that the closest particle to the right
of particle $k$ is labeled $k+1$. If $y_k(t)$ denotes the position of
particle $k$ at time $t$, the configuration of the particle system at
time $t$ is described by the increasing bisequence
$\{y_k(t)\}_{k=-\infty}^{\infty}$.  We can then specify how the
position of the particle at the time step $t+1$ depends on positions
of the particle and its neighbours at the time step $t$. For example,
for rule 184 one obtains
\begin{equation} \label{r184-00}
y_k(t+1)= y_n(t) + \min\{y_{k+1}(t) - y_k(t)-1,1\}.
\end{equation}
Equation (\ref{r184-00}) is sometimes referred to as the motion
representation. The motion representation is analogous to Lagrange
representation of the fluid flow, in which we observe individual
particles and follow their trajectories \cite{MatsukidairaN03}.
It turns out that  the motion representation can be constructed for 
arbitrary number-conserving CA rule by employing algorithm described
in \cite{paper10}.  

\section{Second-order invariants}
In what follows, we will referr to the number of variables of $\xi$ as the order
of the invariant, equal to $\alpha+1$. 
Since the invariant of $\alpha=0$ and corresponding fundamental diagrams have been extensively
studied, we will explore the case of $\alpha=1$, i.e., second order invariants,
using the method of  \cite{Hattori91}.

The arguments $x_0,x_1$ of the density function take values in the set $\{0,1\}$, and therefore $\xi$ can be defined 
in terms of four parameters
\begin{equation}
\xi(0,0)=c_{00}, \,\,
\xi(0,1)=c_{01}, \,\,
\xi(1,0)=c_{10}, \,\,
\xi(1,1)=c_{11}, 
\end{equation} 
where $c_{00},c_{01},c_{10},c_{11} \in \mathbbm{R}$. This can be also expressed as
\begin{eqnarray*}
\xi(x_0,x_1)=c_{00} (1-x_0)(1-x_1)+
c_{01} (1-x_0) x_1+
c_{10} x_0 (1-x_1)+
c_{11} x_0 x_1\\
=c_{00} + (c_{10}-c_{00})x_0 + (c_{01}-c_{00})x_1 + (c_{00}-c_{01}-c_{10}+c_{11})x_0 x_1. 
\end{eqnarray*} 
The constant term does not bring anything new, so we can set $c_{00}=0$. Moreover,
note that for any function $g(x)$ and any $x_0,x_1,\ldots,x_{L-1}\in \{0,1\}$
we have
\begin{equation}
\sum_{i=0}^{L-1}\xi(x_i,x_{i+1})=\sum_{i=0}^{L-1}\big(\xi(x_i,x_{i+1}) + g(x_i) - g(x_{i+1})\big	),
\end{equation} 
which means that if $\xi(x_0,x_1)$ is a density function of some conserved additive quantity,
then $\widehat{F}(x_0,x_1)=\xi(x_0,x_1)+g(x_0)-g(x_1)$ is also a density function of a conserved additive 
quantity. To remove this ambiguity, we will require that $\xi(0,x_1)=0$, similarly as done in~\cite{Hattori91}.
This yields $c_{01}=0$, and we are left with $\xi$ depending on two parameters only
\begin{equation}
\xi(x_0,x_1)=c_{10}x_0  + (c_{11}-c_{10})x_0 x_1.
\end{equation}  
Defining $a_1=-c_{10}$, $a_2=c_{11}-c_{10}$, we arrive at the final parameterization of $\xi$
\begin{equation}
\xi(x_0,x_1)=a_1 x_0  + a_2 x_0 x_1, \mbox{\,\,\,} a_1,a_2 \in  \mathbbm{R}.
\end{equation}

For $\alpha=1$, $l=-1$, and $n=3$, eq. (\ref{gencondition}) becomes
\begin{equation} \label{seconordcond}
 G ( x_0, x_1, x_2,x_3) - \xi(x_{1},x_2) = J ( x_0, x_1, x_2)- J (x_1, x_2, x_3),
 \end{equation}
where
\begin{eqnarray*}
G ( x_0, x_1, x_2,x_3)&=&\xi(f( x_0, x_1, x_2), f(x_1, x_2,x_3))\\
J ( x_0, x_1, x_2) &=& -G(0,0,0,x_0)-G(0,0,x_0,x_1) - G(0,x_0,x_1,x_2)\\
&&\mbox{\,}+\xi(0,0)+\xi(0,x_0)+\xi(x_0,x_1).
\end{eqnarray*} 
Since $\xi(0,0)=\xi(0,x_0)=0$, the formula for current simplifies to
\begin{equation}
J ( x_0, x_1, x_2) = -G(0,0,0,x_0)-G(0,0,x_0,x_1) - G(0,x_0,x_1,x_2) +\xi(x_0,x_1).
\end{equation} 
For a given elementary CA rule $f(x_0,x_1,x_2)$, one can write  eq. (\ref{seconordcond}) 
for all $2^4$ combinations of values of the variables $x_0,x_1,x_2,x_3 \in \{0,1\}$, thus obtaining
an overdetermined linear system of 16 equations with two unknowns $a_1,a_2$.
This system is homogeneous, therefore the solution, if it exists, is not unique.
That is, if $(a_1,a_2)$ is a solution, then $(c a_1, c a_2)$ is also a solution for any $c \in  \mathbbm{R}$.
We will normalize the solution so that the first non-zero number in the pair $(a_1,a_2)$ is
set to be equal to 1.

Solving these equations for all ``minimal'' CA rules\footnote{Elementary CA rules fall into 
88 equivalence classes with respect to the group of transformations generated by the spatial
reflection and the Boolean conjugacy. Minimally-numbered element of each class  are known as
``minimal rules''.}, one finds that for most CA rules
solutions do not exist. Remaining CA rules can be divided into two classes.
The first class contains rules 204 and 170, and for these rules, any pair $(a_1,a_2)$
is a solution. We will not be concerned with these rules, since they exhibit trivial dynamics.
The second class consists of 10 rules for which a unique solution exists (up to the normalization described earlier).
These rules are 12, 14, 15, 34, 35, 42, 43, 51, 140, 142, and 200, as reported in  \cite{Hattori91}.
 
Table 1 shows the density function $\xi$ and  the current $J$ for all of them. The formulas for the current
have been obtained using the HCELL C++ library for cellular automata developed by the author.
\begin{table}
\begin{center}
\begin{tabular}{|c|c|c|}\hline
Rule number & $\xi(x_0,x_1,x_2)$ & $J(x_0,x_1)$ \\ \hline
12  &  $ x_0-x_0 x_1$   & $ -x_0 x_1$ \\ \hline
14  &  $ x_0-x_0 x_1$   & $ -x_0 x_1$ \\ \hline
15  &  $ x_0-x_0 x_1$   & $ -x_0 x_1$ \\ \hline
34  &  $ x_0-x_0 x_1$   & $ -x_1$ \\ \hline
35  &  $ x_0-x_0 x_1$   & $ -x_1$ \\ \hline
42  &  $ x_0-x_0 x_1$   & $ -x_1+x_1 x_2-x_0 x_1 x_2$ \\ \hline
43  &  $ x_0-x_0 x_1$   & $ -x_1+x_1 x_2-x_0 x_1 x_2$ \\ \hline
51  &  $ x_0-x_0 x_1$   & $ -x_1$ \\ \hline
140  &  $ x_0-x_0 x_1$  & $ -x_0 x_1+x_0 x_1 x_2$ \\ \hline
142  &  $ x_0-x_0 x_1$  & $ -x_0 x_1+x_0 x_1 x_2$ \\ \hline
200  &  $ x_0 x_1 $     & $ 0$ \\ \hline
\end{tabular}
\end{center}
\caption{Density of the invariant $\xi$ and the current $J$ for all non-trivial elementary
CA with second order invariants.}
\end{table}  

\section{Fundamental diagrams}
In order to construct fundamental diagrams for rules of Table 1, we first note that
the current for rule 200 is identically equal to zero, thus the equilibrium current
$j(\rho, \infty)=0$. The graph of $j(\rho, \infty)$ vs. $\rho$ for this rule is, therefore,
not interesting.

For all other rules, the density of the invariant is given by the same function 
$\xi(x_0,x_1)= x_0-x_0 x_1$. This means that rules 12, 14, 15, 34, 35, 42, 43, 51, 140, and 142
conserve the number of blocks ``10'' in the configuration. In order to construct
their fundamental diagrams, we have to be able to create an initial configuration with a given
number of pairs ``10''. Construction of a configuration of length $L$ with exactly
$m$ pairs ``10'' can proceed according to the following algorithm.
We start with an array of $L$ integers, $\{s_i(0)\}_{i=0}^{L-1}$.  
\begin{enumerate}
\item Set $s_i(0)=0$ for all $i=0,1,\ldots,L-1$.
\item Place the symbol ``C'' at randomly selected site of the array.
Then place another symbol ``C'' at another site randomly selected among all remaining empty
sites. Repeat this procedure until you place exactly $2m$ symbols ``C''.
\item Let $x=0$. Starting from $i=0$, traverse the array filling it with
$x$ values. Every time when you encounter $C$, set $x:=1-x$. Stop when you reach the end
of the array. 
\end{enumerate} 
The average density of the invariant $\xi(x_0,x_1)= x_0-x_0 x_1$ for the configuration obtained
with the above algorithm will be
\begin{equation}
\rho_{av} = \frac{1}{L}\sum_{i=0}^{L-1} \xi(s_i(t),s_{i+1}(t)) =\frac{m}{L},
\end{equation} 
and it will be independent of $t$. We can also define average current at time  $t$
for a configuration with average density $\rho_{av}$ of the invariant  as
\begin{equation}
j_{av}(\rho_{av},t)=\frac{1}{L}\sum_{i=0}^{L-1} J(s_i(t),s_{i+1}(t),s_{i+2}(t)).
\end{equation} 
Graph of $j_{av}(\rho_{av},t)$ vs. $\rho_{av}$ for very large $t$ will approximate the graph
of $j(\rho,\infty)$ (given by eq. \ref{defexpcurrent}) vs. $\rho$, i.e., the fundamental diagram. 

For six rules from Table 1, the fundamental diagram is strictly linear, and the following
expressions for current can be conjectured based on numerical experiments.
\begin{eqnarray*}
\mbox{Rule 12:\,\, }  j(\rho,\infty)&=&0\\
\mbox{Rule 15:\,\, }  j(\rho,\infty)&=&\rho - 1/2\\
\mbox{Rule 34:\,\, }  j(\rho,\infty)&=&-\rho\\
\mbox{Rule 42:\,\, }  j(\rho,\infty)&=&-\rho\\
\mbox{Rule 51:\,\, }  j(\rho,\infty)&=&-1/2\\
\mbox{Rule 140:\,\, }  j(\rho,\infty)&=&0
\end{eqnarray*}

The remaining four rules are more interesting, as they exhibit singularities in fundamental 
diagrams, as shown in Figure 1.
%%%%%%%%%%%%%%%%%%%%%%%%%%%%
\begin{figure}
\begin{center}
\includegraphics[scale=0.55]{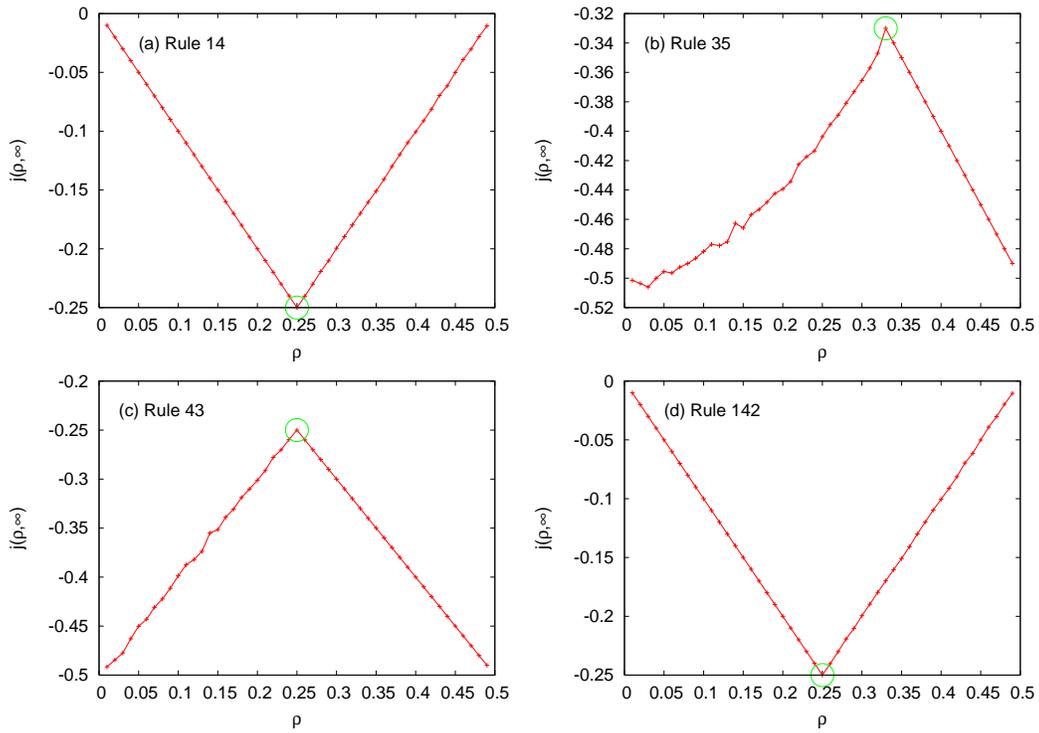}
\end{center}
\caption{Fundamental diagrams for rules 14, 35, 43, 142. Singularities are denoted by green circles.
Diagrams have been obtained using 50000 lattice sites after 50000 iterations.}
\label{fig1}
\end{figure}
%%%%%%%%%%%%%%%%%%%%%%%%%%%%%%
It is remarkable that singularities are present only in fundamental diagram of those particular
rules. In \cite{Hattori91}, authors searched for invariants of up to seventh order for all ``minimal'' 
elementary CA. According to the table published in their paper, rules  $14, 35, 43, 142$ do not
have any other invariant except $\xi(x_0,x_1)=x_0-x_0 x_1$, in contrast to remaining rules of Table 1,
which also posses other higher order invariants. It seems that singularities in the fundamental diagram can appear 
only in rules which have only one invariant, just like rule 184, which only has first order invariant
$\xi(x_0)=x_0$. 

\section{Convergence to equilibrium}
In number-conserving cellular automata, singularities of the fundamental diagram exhibit
critical behavior. This can be illustrated by introducing the the decay time defined as 
\begin{equation} \label{tau}
\tau(\rho) = \sum_{t=0}^{\infty} |j(\rho,t)-j(\rho,\infty)|.
\end{equation}
If the decay of $j(\rho, t)$ toward its equilibrium value $j(\rho,\infty)$ is of power-law type, the above 
sum diverges.
For all rules in Figure 1, we have performed computer simulations
to estimate $\tau$. The value of
$\tau$ has been estimated by measuring $j_{av}(\rho,t)$ for
$t=0,1,\ldots,1000$, and truncating the sum (\ref{tau}) at
$t=1000$. Figure 2 shows a typical graph of $\tau$ as a function of
$\rho$, obtained for rule 42.  Comparing Figures 1c and Figure 2 we clearly see that $\tau$ diverges at
the critical point of rule 42, which occurs at $\rho=0.25$. We will denote this value by $\rho_c$.
%%%%%%%%%%%%%%%%%%%%%%%%%%%%
\begin{figure}
\begin{center}
\includegraphics[scale=1.0]{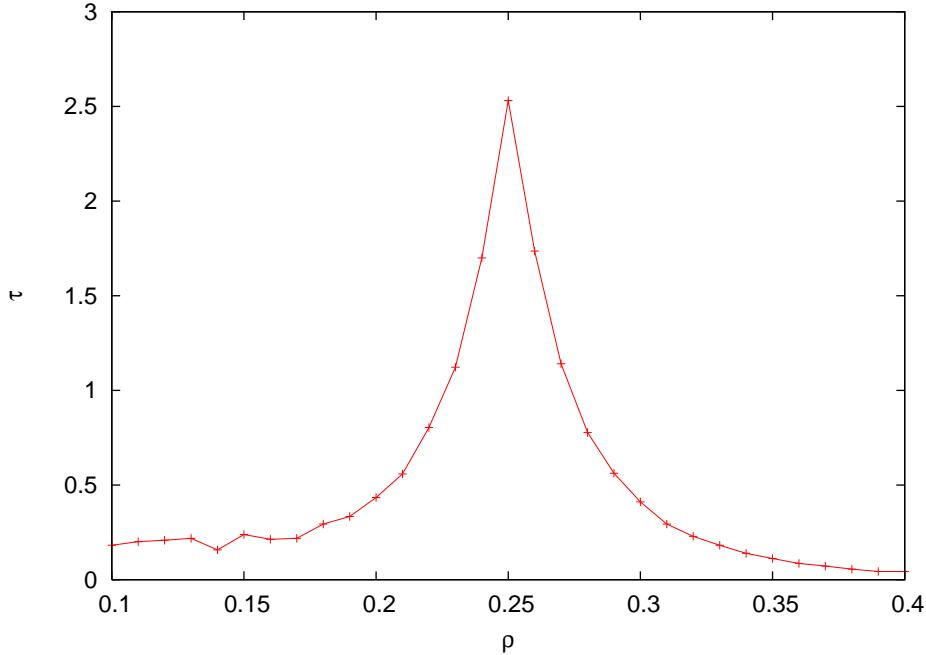}
\end{center}
\caption{Decay time vs. density of the invariant for rule 43.}
\label{fig2}
\end{figure}
%%%%%%%%%%%%%%%%%%%%%%%%%%%%%%
Assuming that $|j(\rho_c,\infty)-j(\rho_c,t)| \sim t^{-\alpha}$, we have
determined the exponent $\alpha$ as the slope of the straight line
which best fits the logarithmic plot of $|j(\rho_c,\infty)-j(\rho_c,t)|$
vs. time $t$. Example of such a plot, again for rule 42, is shown in Figure 3.
Table 2 shows values of the exponent $\alpha$ for critical points of all 
four rules of Figure 1.
%%%%%%%%%%%%%%%%%%%%%%%%%%%%
\begin{figure}
\begin{center}
\includegraphics[scale=1.0]{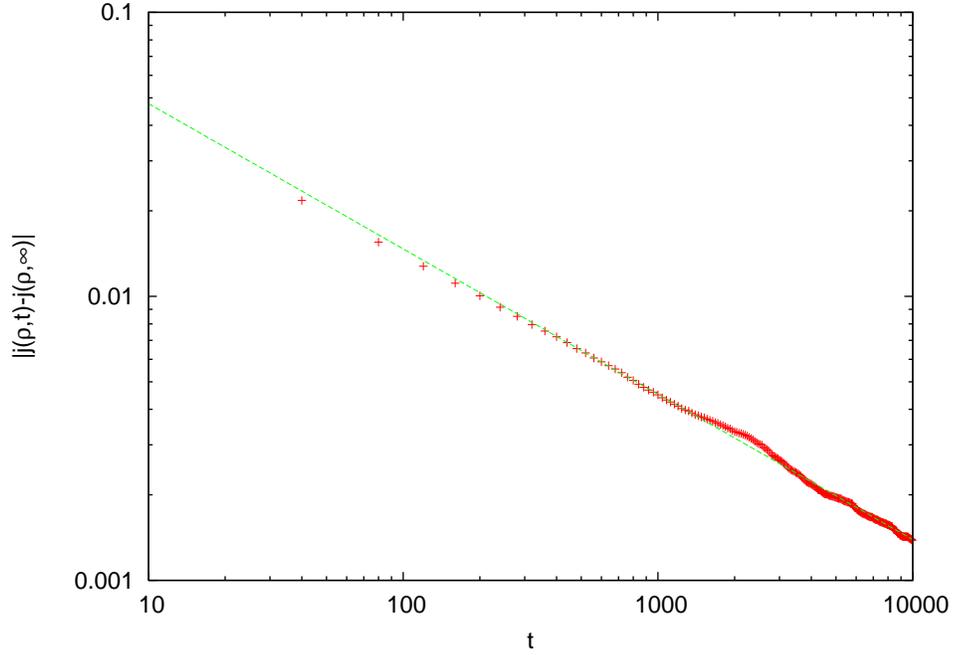}
\end{center}
\caption{Logarithmic plot of $|j(\rho_c,\infty)-j(\rho_c,t)|$ as a function of time for rule 43.
Data points ($+$) represent
computer simulations, while the dashed line represents the best
fit.}
\label{fig3}
\end{figure}
%%%%%%%%%%%%%%%%%%%%%%%%%%%%%%
\begin{table}
\begin{center}
\begin{tabular}{|c|c|c|}\hline
Rule number & $(\rho_c,j(\rho_c,\infty))$ & $\alpha$ \\ \hline
14  & $(1/4,-1/4)$   & $-0.504$ \\ \hline
35  & $(1/3,-1/3)$   & $-0.472$\\ \hline
43  & $(1/4,-1/4)$  & $-0.492$ \\ \hline
142  & $(1/4,-1/4)$  & $-0.502$ \\ \hline
\end{tabular}
\end{center}
\caption{Values of the exponent  $\alpha$  at the critical point for 
elementary CA rules with second-order additive invariant.}
\end{table}  

Exponent $\alpha$ is known to be equal to exactly $1/2$ for rule 184 and its generalizations, and rigorous
proof of this fact exists \cite{paper11}.
Extensive numerical experiments support the conjecture that for all rules  with first-order invariant
 the value $\tau=1/2$ is universal, in the case of both piecewise linear \cite{paper19}   
and nonlinear \cite{paper19} fundamental diagrams. Table 2 provides evidence that
a more general conjecture may be valid: regardless of the order of the invariant,
exponent $\alpha$ seems to have universal value of $1/2$. In the next section we will 
offer some justification for this conjecture for rules with second-order invariants.
\section{Dynamics of localized structures}
In rule 184, the power-law convergence of the current toward its equilibrium value
is related to the dynamics of this rule, which resembles ballistic annihilation.
The spatiotemporal patter generated by rule 184 can be understood as propagation of
two types of localized structures, shown in Figure 4. 
\begin{figure}
\begin{center}
\includegraphics[scale=0.5]{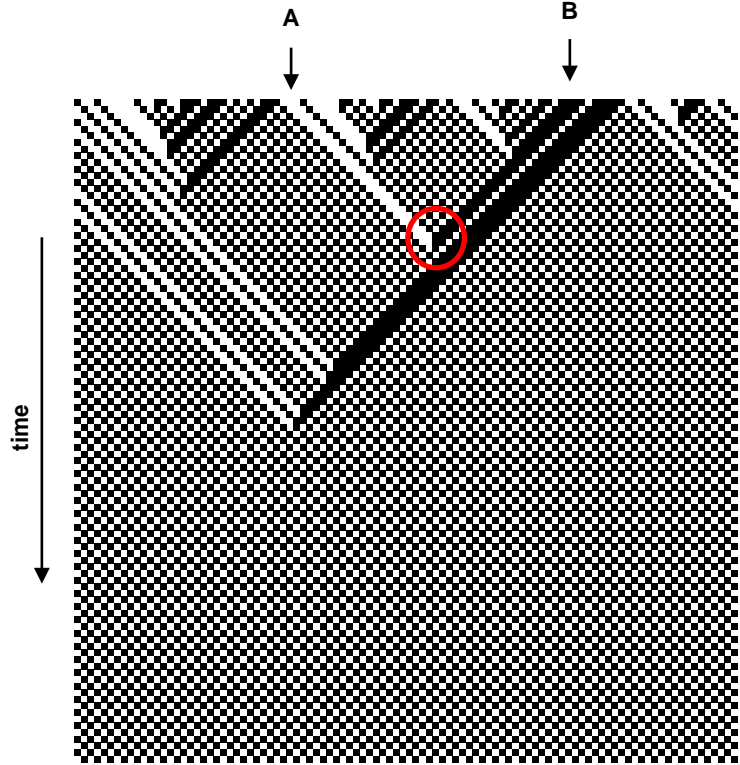}
\end{center}
\caption{Spatiotemporal pattern generated by rule 184. Initial configuration
is represented by the top row, where black squares represent 1's, and white spaces represent 0's.
Consecutive configurations are plotted as consecutive rows. Two types of defects
of ``A'' and ``B'' type are visible, annihilating in the circled spot.}
\label{fig4}
\end{figure}
These two types of structures, marked with letters ``A'' and ``B'', propagate in opposite 
directions and annihilate upon collision. At the critical point, the number of ``A'' defects
in the initial configuration is the same as the number of  ``B'' defects, and it takes long time 
for all of them to disappear, hence the ``critical slowing down'', or power-law convergence is observed. Detailed analysis
of this process \cite{paper11} leads to the exact formula for the current $j(\rho, t)$, which
in the limit of large $t$ and using de Moivre-Laplace limit theorem
leads to  $j(\rho_c,\infty)-j(\rho_c,t) \sim t^{-1/2}$. Here,
by $f(t)\sim g(t)$ we mean that $\lim_{t\to\infty} f(t)/g(t)$
exists and is different from $0$. 

Dynamics of rules 14, 35, 43, 142 resembles rule 184 very strongly, as can be seen
in Figure 5, which shows spatiotemporal patterns at the critical point for all
four rules.
\begin{figure}
\begin{center}
\includegraphics[scale=0.7]{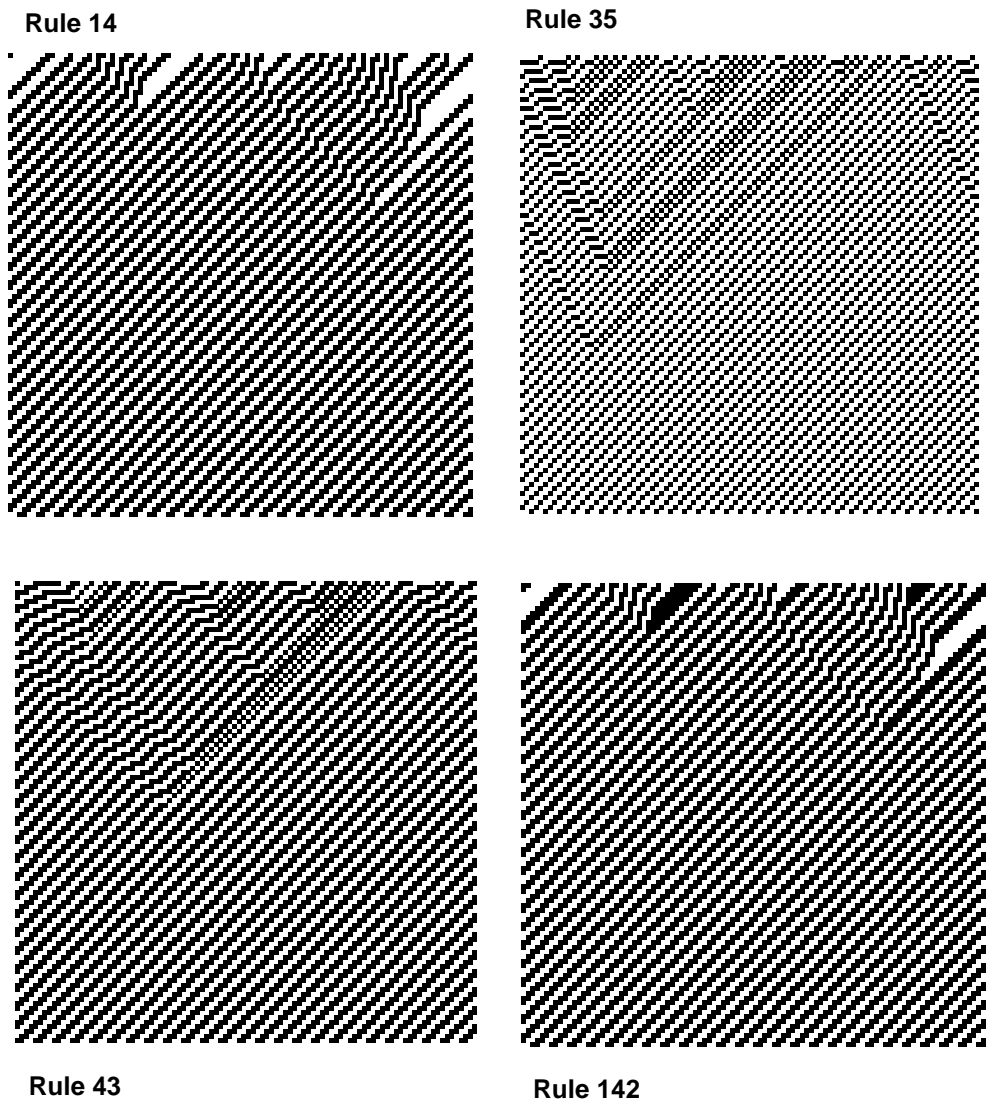}
\end{center}
\caption{Spatiotemporal pattern generated by rules 14, 35, 43, 142
at their critical points.
Density of the invariant $\xi$ equals $1/4$ for rules 14, 43, 142,
and $1/3$ for rule 35. 
}
\label{fig5}
\end{figure}
In all four cases, localized propagating structures moving in opposite directions and annihilating upon collision are visible.
In fact, for two of these rules, it is possible to establish direct relationship with rule 184.
In order to do this, we will define superposition of two rules as
\begin{equation}
(f \circ g) (x_0,x_1,x_2,x_3,x_4)
=f(g(x_0,x_1,x_2),g(x_1,x_2,x_3),g(x_2,x_3,x_4)).
\end{equation} 
If $h \circ f = h \circ g$, then following  \cite{Boccara93} we will say that
$g$ is a transform of rule $f$ by $h$. If by $f_k$ we denote local function of rule $k$,
one can show \cite{Boccara93} that
\begin{eqnarray}
f_{60} \circ f_{43} = f_{184} \circ f_{60},\\
f_{60} \circ f_{142} = f_{226} \circ f_{60}.
\end{eqnarray} 
This means that there exists a local mapping (rule 60) which transforms rule 43 into rule 184, and rule 142 into rule 226 (recall that rule 226 is 
the image of rule 184 under spatial reflection). Similarity of dynamics of rules 43, 142 to rule 184
is, therefore, not surprising. 

\section{Conclusion}
We investigated second order additive invariants in elementary cellular automata rules. We found that 
fundamental diagrams of rules which possess additive invariant are either linear or exhibit
singularities similar to singularities of rules with first-order invariant. Singularities can appear only
in rules with exactly one invariant. At the critical
density of the invariant, the current decays to its equilibrium value as a  power law $t^{\alpha}$,
and  the value of the exponent $\alpha$ obtained from numerical simulations is very close
to $-1/2$. This indicates that regardless of the order of the invariant, the dynamics of 
CA rules with invariants is very similar. 

Since rules 43 and 142 can be transformed into rules 184 and 226 by a surjective local transformation,
it should be possible to obtain for them rigorous formulas for the expected value of the current at
arbitrary time, similarly as it has been done for rule 184 and its generalizations \cite{paper11}. Such formula
could then used to compute the exact value of the exponent $\alpha$. For rules 14 and 35 no such local transformation
exists, nevertheless they exhibit localized propagating structures strikingly similar to 
structures of rule 184, so exact calculation of the current might be possible too. This problem is
currently under investigation and will be reported elsewhere.

 \vskip 1cm
 \noindent \textbf{Acknowledgements:} The author
acknowledges financial support from NSERC (Natural Sciences and
Engineering Research Council of Canada) in the form of the
Discovery Grant.

\providecommand{\href}[2]{#2}\begingroup\raggedright\endgroup

\end{document}